\documentclass[]{spie}  %>>> use for US letter paper
\usepackage[]{graphicx}
\usepackage{hyperref}

\title{The hunt for Sirius \emph{\LARGE Ab}: Comparison of algorithmic sky and PSF estimation performance in deep coronagraphic thermal-IR high contrast imaging}

\author{Joseph D. Long\supit{a}, Jared R. Males\supit{a}, Katie M. Morzinski\supit{a}, Laird M. Close\supit{a}, Frans Snik\supit{b}, Matthew A. Kenworthy\supit{b}, Gilles P. P. L. Otten\supit{b}, John Monnier\supit{c}, Volker Tolls\supit{d}, \\and Alycia Weinberger\supit{e}
\skiplinehalf
\supit{a}Steward Observatory, University of Arizona, Tucson, 933 N Cherry Ave.,
Tucson, AZ 85721, USA;
\supit{b}Leiden Observatory, Leiden University, P.O. Box 9513, 2300 RA Leiden, The Netherlands;
\supit{c}University of Michigan, Department of Astronomy, 1085 S. University, Ann Arbor, MI 48109, USA;
\supit{d}Harvard-Smithsonian Center for Astrophysics, 60 Garden St., Cambridge, MA 02138, USA;
\supit{e}Carnegie Institution of Washington, Department of Terrestrial Magnetism, 5241 Broad Branch Road, N.W., Washington, D.C. 20015, USA}
\newcommand{\micron}{\ensuremath{\mu \mathrm{m}} }

\begin{document}
  \maketitle

%%%%%%%%%%%%%%%%%%%%%%%%%%%%%%%%%%%%%%%%%%%%%%%%%%%%%%%%%%%%%
\begin{abstract}
Despite promising astrometric signals, to date there has been no success in direct imaging of a hypothesized third member of the Sirius system. Using the Clio instrument and MagAO adaptive optics system on the Magellan Clay 6.5 m telescope, we have obtained extensive imagery of Sirius through a vector apodizing phase plate (vAPP) coronagraph in a narrowband filter at 3.9 microns. The vAPP coronagraph and MagAO allow us to be sensitive to planets much less massive than the limits set by previous non-detections. However, analysis of these data presents challenges due to the target's brightness and unique characteristics of the instrument. We present a comparison of dimensionality reduction techniques to construct background illumination maps for the whole detector using the areas of the detector that are not dominated by starlight. Additionally, we describe a procedure for sub-pixel alignment of vAPP data using a physical-optics-based model of the coronagraphic PSF.
\end{abstract}

%>>>> Include a list of keywords after the abstract

\keywords{thermal infrared, background subtraction, high-contrast imaging}

%%%%%%%%%%%%%%%%%%%%%%%%%%%%%%%%%%%%%%%%%%%%%%%%%%%%%%%%%%%%%
\section{INTRODUCTION}
\label{sec:intro}  % \label{} allows reference to this section

Sirius, in Canis Major, is the brightest star visible from Earth in terms of apparent brightness. Unsurprisingly, it is a well studied system---astrometric measurements suggested\cite{Bessel1844} and later direct imaging confirmed the existence of a companion star Sirius B perturbing the orbit of Sirius A as long ago as 1862\cite{Bond1862}. Thanks to its proximity and brightness, Sirius remains a compelling target for exoplanet searches in the present day. Thalmann {\it et al.} conducted observations with the Subaru and MMT telescopes to establish limits on the separation and mass of any companions thus far undetected, robustly refuting the existence of companions in the 6-12 $M_\mathrm{jup}$ range at a separation of $1''$\cite{Thalmann2011}. As we achieve greater contrasts and smaller inner working angles, the system is worth revisiting in search of lower-mass companions to Sirius A at smaller separations.

To that end, the authors (led by JRM) obtained deep imagery of Sirius A with the Magellan Clay 6.5-meter telescope in a 2015 observing campaign. The data were taken with MagAO\cite{Close2012} and the Clio instrument with a vector-apodizing phase plate coronagraph (vAPP)\cite{Otten2017} in 500 millisecond integrations, totaling 107 minutes of integration in a 3.9 \micron narrowband filter.

The extreme brightness of Sirius led to challenges in both data acquisition and analysis, requiring the development of new strategies to handle saturated pixels and thermal background fluctuations. In this paper, we describe strategies to remove instrumental and thermal background signal from the data in order to achieve the best possible contrast, as well as a procedure we developed to enable the subpixel alignment of vAPP coronagraph PSFs for coronagraphic post-processing.

\section{BACKGROUND}
\label{sec:background}

The data were taken in the 3.9 \micron narrowband filter on Clio through a vAPP coronagraph\cite{Otten2017}. In this bandpass, thermal emission from the atmosphere and telescope structure adds a time-varying background signal to our measurements. The conventional strategy for background estimation in thermal infrared imaging involves ``nodding'' the pointing of the telescope to place the image of the star at different locations on the detector. Images with the star at one location are then calibrated against background levels estimated from the same region of the detector in subsequent images where the star's image has been moved.

Sirius, due to its extreme brightness, was observed at a single detector location with background frames taken approximately every twenty minutes. To obtain background measurements, the telescope pointing was changed to move the star entirely off-detector.
% nodded off-chip because sirius is so bright that the only way to avoid getting airy rings on detector is to go several arcminutes away, not just across the detector
Eight blocks of 160 sky frames each were taken to measure the background level and monitor its variation over the course of the night. The mean level varies by hundreds of counts between blocks, meaning that to achieve the best subtraction we will have to model the variation over the course of our science exposures by measuring background regions in the science frames.

Additionally, the Sirius PSF saturates the Clio detector even in 500 millisecond integrations. Subpixel-accurate PSF locations are a prerequisite for aligning the frames, which in turn is a prerequisite for coronagraphic image post-processing (e.g. KLIP\cite{Soummer2012}). The technique we use to align images is based on a Fourier transform cross-correlation, which meant we had to develop a realistic model for the PSF as input. The alignment process and development of models for the PSF are detailed in Section~\ref{sec:psf-refinement}.

\section{PSF DETECTION}
\label{sec:detection}

%-------------
\begin{figure}
  \begin{center}
  \begin{tabular}{c}
  \includegraphics{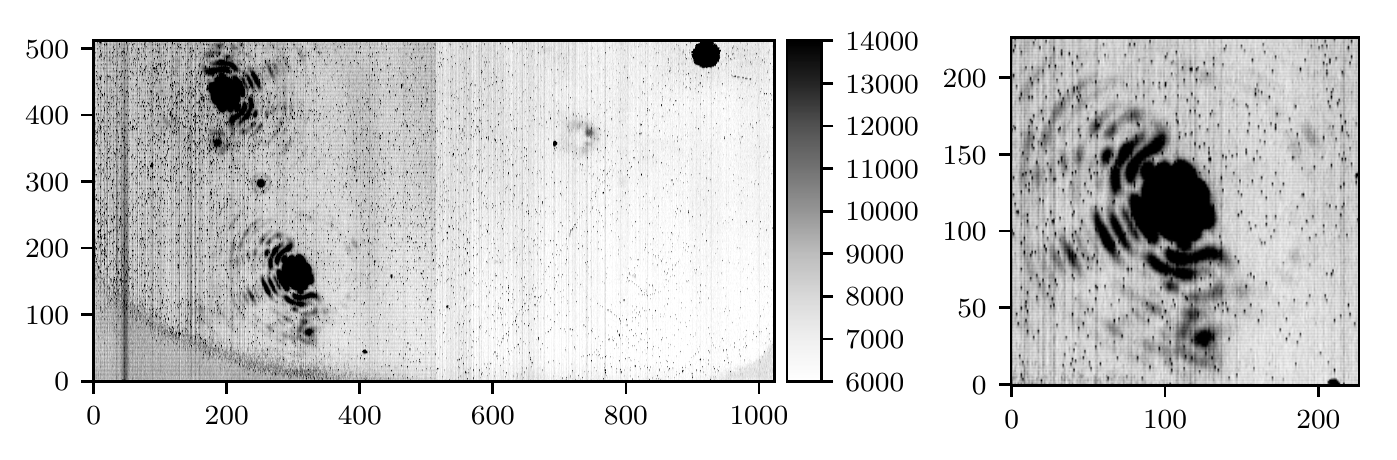}
  \end{tabular}
  \end{center}
  \caption[raw-image-detail]
%>>>> use \label inside caption to get Fig. number with \ref{}
  { \label{fig:raw-image-detail} The first panel shows a raw Clio+vAPP Sirius frame before any corrections have been applied.
The second panel shows a $227 \times 227$ cutout centered on the lower vAPP PSF. Both panels are scaled linearly from the 5th to 98th percentile pixel value in the raw frame.}
  \end{figure}
%-------------

%raw_Sirius2_split_011656

Simple peak-finding algorithms based on identifying the highest-valued pixels were stymied by the irregular background illumination pattern and the saturation of the peak, which led to a broad and flat core of the PSF and first Airy ring (Figure~\ref{fig:raw-image-detail}).

\begin{table}[h]
  \caption{The regions that our PSF detection algorithm searches are mostly defined relative to the middle or leakage PSF in the three-part vAPP PSF arrangement. Initial guesses for the leakage PSF locations are hard-coded based on the two beam nodding locations defined for Clio vAPP observations.}
  \label{tab:search-regions}
  \begin{center}
  \begin{tabular}{|l|l|l|} %% this creates two columns
  %% |l|l| to left justify each column entry
  %% |c|c| to center each column entry
  %% use of \rule[]{}{} below opens up each row
  \hline
  \rule[-1ex]{0pt}{3.5ex}  {\bf Region name} & {\bf Initial location} & {\bf Search box shape} \\
  \hline
  \rule[-1ex]{0pt}{3.5ex}  Leakage PSF (left nod location) & $(x_{1}, y_{1}) = (250, 287)$ & $75 \times 75$ \\
  \hline
  \rule[-1ex]{0pt}{3.5ex}  Top PSF (left) & $(x_{1} - 60, y_{1} + 130 )$ & $227 \times 227$\\
  \hline
  \rule[-1ex]{0pt}{3.5ex}  Bottom PSF (left) & $(x_{1} + 60, y_{1} - 130 )$ & $227 \times 227$\\
  \hline
  \rule[-1ex]{0pt}{3.5ex}  Leakage PSF (right nod location) & $(x_{2}, y_{2}) = (630, 311)$ & $75 \times 75$ \\
  \hline
  \rule[-1ex]{0pt}{3.5ex}  Top PSF (right) & $(x_{2} - 60, y_{2} + 130)$ & $227 \times 227$\\
  \hline
  \rule[-1ex]{0pt}{3.5ex}  Bottom PSF (right) & $(x_{2} + 60, y_{2} - 130 )$ & $227 \times 227$\\
  \hline
  \rule[-1ex]{0pt}{3.5ex}  Stray light ``ghost'', upper right & $(800, 350)$ & $350 \times 150$\\
  \hline
  \rule[-1ex]{0pt}{3.5ex}  Stray light ``glint'', lower left & $(130, 95)$ & $90 \times 90$\\
  \hline
  \end{tabular}
  \end{center}
  \end{table}

Fortunately, for the purposes of peak finding, a simple median across the sky estimation frames combined with linear scaling to match the median pixel values provided a good enough estimate of the background illumination pattern. As discussed in the following section, very little of the detector in a science frame is actually background signal. This means that this rough background estimate leads to oversubtraction, which fortunately does not affect our ability to use the frame for peak finding. The raw frame and median sky are supplied to the rough PSF location function, which subtracts a scaled median sky frame from the raw frame before convolving with a 15 pixel Gaussian kernel. The rough peaks are then determined by the relative maxima in search regions defined by the two nod locations for Clio vAPP data. For these data, the search regions were defined according to Table \ref{tab:search-regions}.

The rough PSF location function returns an approximate peak location (as integer pixel coordinates), as well as a boolean value indicating whether the location was trustworthy (i.e. whether a peak was found). Certain glints within Clio depend on various instrument settings and the telescope orientation, and are not found within every frame. Additionally, both nod positions are searched for peaks, so at least three of every eight regions evaluated will not contain the peaks we are looking for.

The heuristic we decided on to identify empty search regions was that the maxima were located at the edges of the search region rather than near the center (i.e. the only signal was a smooth gradient from oversubtracted background). There are certainly more sophisticated ways to do this, but this technique worked surprisingly well.

\section{SKY BACKGROUND ESTIMATION}
\label{sec:background-estimation}

The background levels during our sky observations varied from $\approx 7670$ to $\approx 7880$ counts over the course of the night. Of course, the true background level for every pixel during a science frame cannot be observed, so we must necessarily estimate the level and pattern of background illumination. We are able to do this using the values of the pixels of the science frames that are not dominated by the star's diffraction pattern or stray light.

Inspired by the amazing utility of low-rank approximation methods (e.g. KLIP) in PSF reconstruction, we undertook a reconstruction of the full background from the regions of the detector that were not dominated by starlight. By constructing a basis of eigenimages from the first six components of the PCA of the sky frames, we were able to reconstruct the sky background in a withheld test set of sky frames to better than 14 counts RMS error.

Of course, reconstructing the sky illumination using a complete sky frame is a different problem from reconstructing the sky illumination using only those parts of a science frame that are not dominated by starlight. The Clio instrument contains several surprising glints and artifacts (described in Ref.~\citenum{Morzinski2015} and Ref.~\citenum{Otten2017}), some of which are only visible after background subtraction. Since Sirius is the brightest star in the sky, a number of negligible glints become nonnegligible when the instrument is pointed at it. Additionally, effects that scale with flux, such as the negative image created 512 pixels from the stellar PSF due to amplifier crosstalk discussed in Ref.~\citenum{Morzinski2015}, are more apparent.

%-------------
\begin{figure}
  \begin{center}
  \begin{tabular}{c}
  \includegraphics{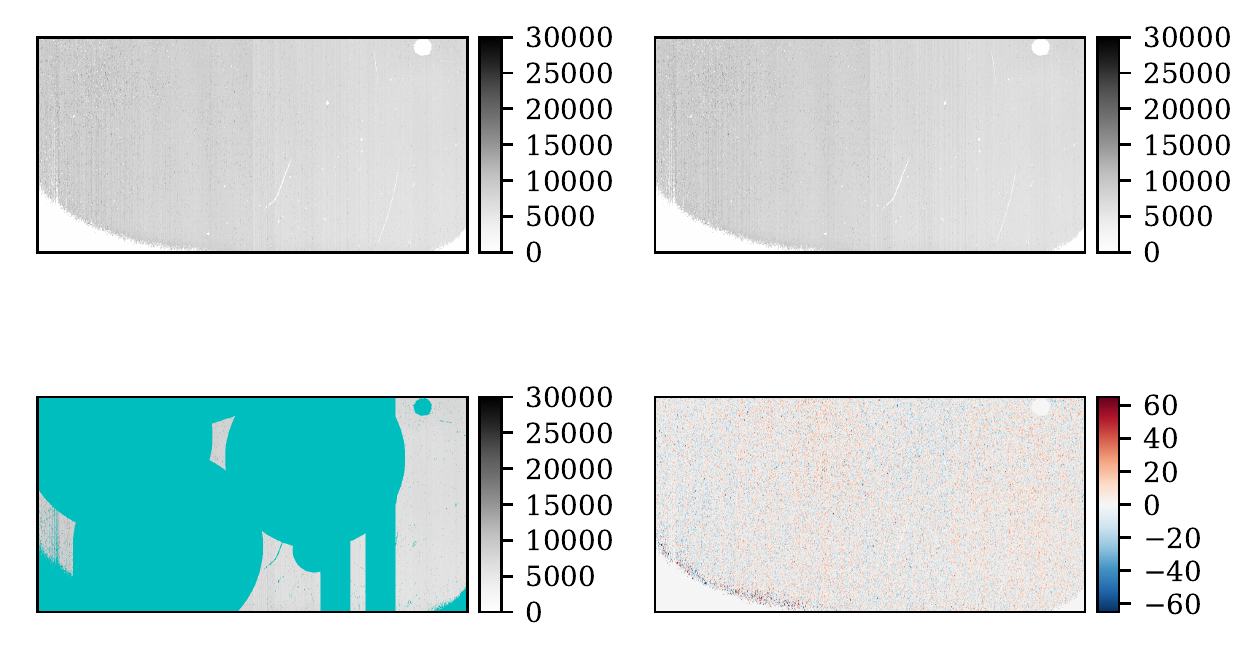}
  \end{tabular}
  \end{center}
  \caption[characterize-background]
%>>>> use \label inside caption to get Fig. number with \ref{}
  { \label{fig:characterize-background}
Upper left: a single frame from the sky backgrounds taken as part of these observations that was not used to compute the PCA basis. Upper right: a reconstructed background illumination pattern from a 6 component PCA basis. Lower left: union of bad pixel mask, instrumental glints mask, and science PSF masks overlaid in cyan on the frame from upper left. Lower right: Residuals from subtracting upper left from upper right image, scaled to $\pm$99.9\%ile values.}
  \end{figure}
%-------------

\begin{table}[h]
  \caption{The regions we exclude from the background reconstruction. The rough location of the top coronagaphic PSF is denoted $x_\mathrm{top}, y_\mathrm{top}$, the bottom $x_\mathrm{bottom}, y_\mathrm{bottom}$, the upper-right ``ghost'' $x_\mathrm{ghost}, y_\mathrm{ghost}$, and the lower-left ``glint'' $x_\mathrm{glint}, y_\mathrm{glint}$. Regions defined relative to the glint or ghost locations are not masked if no peak was found in the corresponding region in the PSF location step. These relationships only hold for the left nod location.}
  \label{tab:mask-regions}
  \begin{center}
  \begin{tabular}{|l|l|l|} %% this creates two columns
  %% |l|l| to left justify each column entry
  %% |c|c| to center each column entry
  %% use of \rule[]{}{} below opens up each row
  \hline
  \rule[-1ex]{0pt}{3.5ex}
    {\bf Region name} &
    {\bf Center} &
    {\bf Mask shape} \\
  \hline
  \rule[-1ex]{0pt}{3.5ex}
    Top PSF &
    $x_\mathrm{top}, y_\mathrm{top}$ &
    circular, $r = 226$ px \\
  \hline
  \rule[-1ex]{0pt}{3.5ex}
    Top PSF spike &
    $x_\mathrm{top}, y_\mathrm{top}$ &
    rectangular, $900 \times 82$ pixels, rotated 27$^\circ$ CCW\\
  \hline
  \rule[-1ex]{0pt}{3.5ex}
    Bottom PSF &
    $x_\mathrm{bottom}, y_\mathrm{bottom}$ &
    circular, $r = 226$ px \\
  \hline
  \rule[-1ex]{0pt}{3.5ex}
    Bottom PSF spike &
    $x_\mathrm{bottom}, y_\mathrm{bottom}$ &
    rectangular, $900 \times 82$ px, rotated 27$^\circ$ CCW\\
  \hline
  \rule[-1ex]{0pt}{3.5ex}
    Stray light blob 1 &
    $x_\mathrm{top} + 470, y_\mathrm{top} - 51$ &
    circular, $r = 214$ px \\
  \hline
  \rule[-1ex]{0pt}{3.5ex}
    Stray light blob 2 &
    $x_\mathrm{top} + 468, y_\mathrm{top} - 269$ &
    circular, $r = 52$ px \\
  \hline
  \rule[-1ex]{0pt}{3.5ex}
    Stray light blob 3 &
    $x_\mathrm{top} + 595, y_\mathrm{top} - 137$ &
    circular, $r = 70$ px \\
  \hline
  \rule[-1ex]{0pt}{3.5ex}
    Stray light ``ghost'' &
    $x_\mathrm{ghost}, y_\mathrm{ghost}$ &
    circular, $r = 66$ px \\
  \hline
  \rule[-1ex]{0pt}{3.5ex}
    Stray light ``ghost'' twin &
    $x_\mathrm{ghost} - 63, y_\mathrm{ghost} + 179$ &
    circular, $r = 66$ px \\
  \hline
  \rule[-1ex]{0pt}{3.5ex}
    Stray light ``glint'' &
    $x_\mathrm{glint}, y_\mathrm{glint}$ &
    square, $90 \times 90$ px \\
  \hline
  \rule[-1ex]{0pt}{3.5ex}
    Negative top PSF image &
    $x_\mathrm{top} + 512, 256$ &
    rectangular, $71 \times 512$ px \\
  \hline
  \rule[-1ex]{0pt}{3.5ex}
    Negative bottom PSF image &
    $x_\mathrm{bottom} + 512, 256$ &
    rectangular, $71 \times 512$ px \\
  \hline
  \end{tabular}
  \end{center}
\end{table}

%-------------
\begin{figure}
  \begin{center}
  \begin{tabular}{c}
  \includegraphics{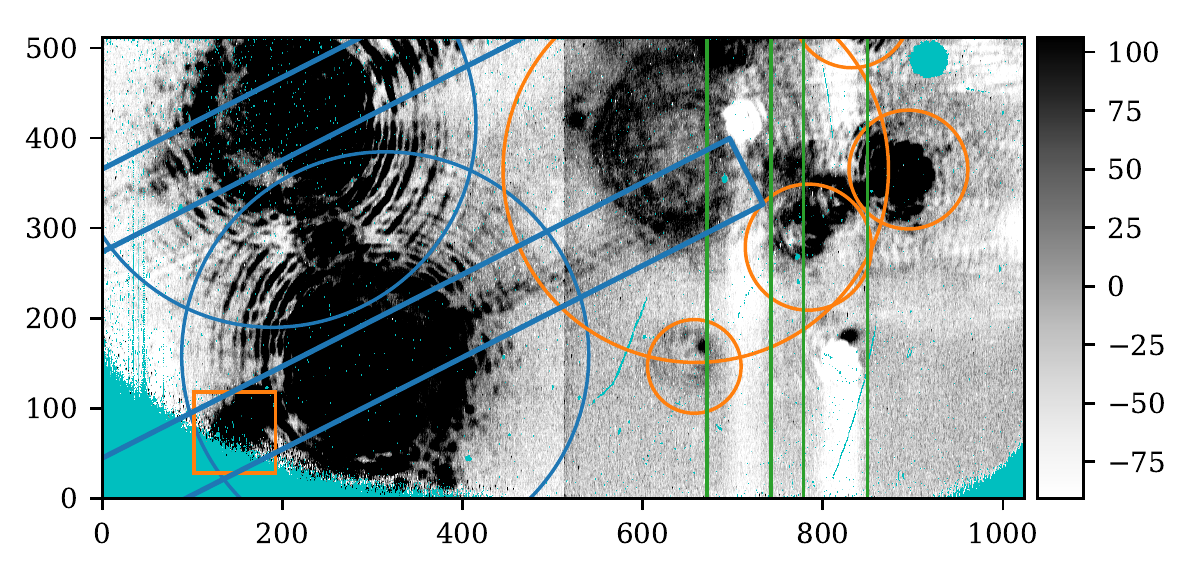}
  \end{tabular}
  \end{center}
  \caption[mask-shapes]
%>>>> use \label inside caption to get Fig. number with \ref{}
  { \label{fig:mask-shapes}
A calibrated and background-subtracted frame from the Sirius dataset with masked shapes overlaid. Bad pixels are shown in cyan, and the values have been scaled linearly from 5th \%ile to 80th \%ile to show faint structure. Science PSF structure masks are outlined in blue. Instrumental stray light mask shapes are outlined in orange. The columns containing negative PSF images due to amplifier crosstalk are masked as well, and shown in green. Note that although some mask shapes partially overlap others in this figure, stray light artifacts may move independently of each other over the course of the observations.}
  \end{figure}
%-------------

To be as conservative as possible, while preserving as much background signal as possible for reconstruction, we created a function to produce a mask for each frame based on the rough PSF locations detected in the previous step. The shapes that make up the mask are shown overlaid on a background-subtracted Sirius frame in Figure~\ref{fig:mask-shapes}, and the centers, sizes, and shapes are detailed in Table~\ref{tab:mask-regions}.

When masking out this much of the detector, reconstructing complete sky frames becomes more complicated. For example, the length of the vectors used to construct the PCA basis and the vectors made from the unmasked pixels of the image are of different length. We experimented with various imputation strategies for the missing values, including replacement with values from a median sky frame. Ultimately, the best way to minimize reconstruction error was to treat it as a least-squares problem to find the coefficients for each component (eigenimage) that minimized the RMS error when comparing only the unmasked pixels. This is the same approach as that taken by Ref.~\citenum{Hunziker2018}, which was published while this work was ongoing.

Using the same test set of sky frames withheld from the PCA computation and masking out regions based on PSF locations from a science frame shows that we can reconstruct the sky frames in the test set to 13.5 counts RMS error overall (12.9 counts in the pixels used for estimation, 13.8 counts over the rest of the pixels) as shown in Figure~\ref{fig:characterize-background}. The thermal background thus subtracted (to within better than a percent of its true value, based on our cross-validation), we are able to perform a more precise measurement of the PSF centers.

\section{PSF LOCATION REFINEMENT}
\label{sec:psf-refinement}

Estimating the background intensity from the science frames depends on knowing the PSF location, as well as the location of any internal glints. However, while the mask generation function for background estimation is robust to small inaccuracies in the PSF locations, creating a set of images aligned at the sub-pixel level requires more precise knowledge of the PSF locations.

Precise PSF locations, in turn, are hard to determine without a model of the background. So, our pipeline uses the rough PSF centers and background-subtracted frames as input to a location refinement algorithm. The core of the algorithm consists of the following steps:

%-------------
\begin{enumerate}
  \item Cut out a square region of the background-subtracted frame centered on the initial, rough location for the top vAPP PSF $(x, y)$ (filling in any out-of-bounds regions or bad pixels with the minimum value)
  \item Compute a Fourier transform cross-correlation between the PSF cutout and an identically sized model for the PSF (e.g. from a physical optics simulation)
  \item Find the displacement $(\Delta x, \Delta y)$ from the cross-correlation peak
  \item Repeat steps 1-3 for the bottom, and leakage PSFs and find the average displacement $(\overline{\Delta x}, \overline{\Delta y})$ for all three
  \item Compute new center coordinates $(x', y') = (x + \overline{\Delta x}, y + \overline{\Delta y})$ for the top, bottom, and leakage PSFs.
\end{enumerate}

%-------------
\begin{figure}
  \begin{center}
  \begin{tabular}{c}
  \includegraphics{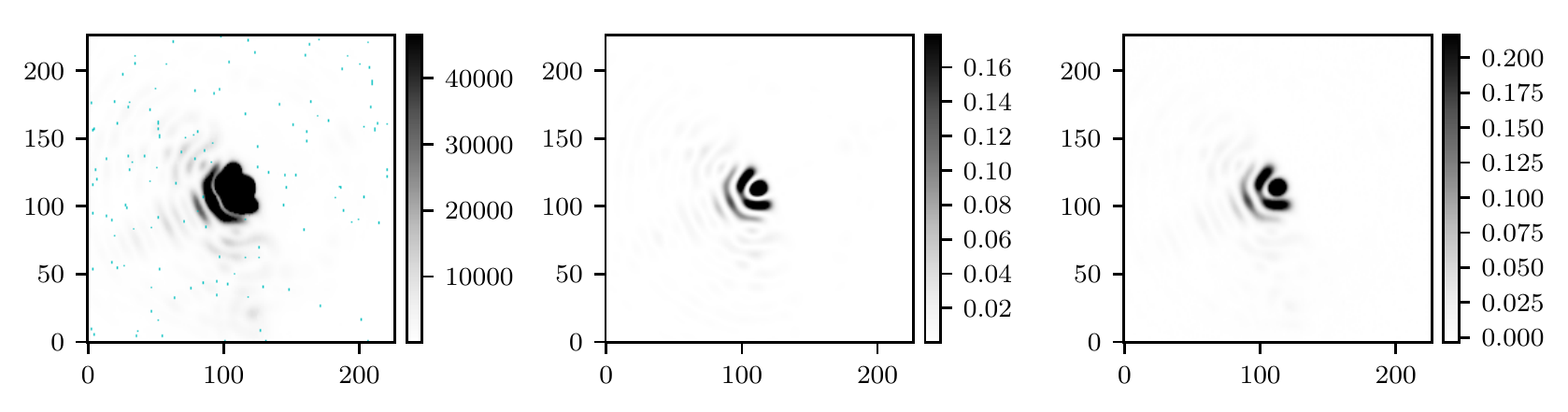}
  \end{tabular}
  \end{center}
  \caption[psf-model-comparison]
%>>>> use \label inside caption to get Fig. number with \ref{}
  { \label{fig:psf-model-comparison}
The first panel shows a $227 \times 227$ cutout from a science frame of Sirius through Clio+vAPP that has been corrected for non-linearity of detector response and background-subtracted. (Bad pixels are flagged in cyan.) The core and first two Airy rings have saturated, and blurring has filled in the gaps. The second panel shows a POPPY model for the vAPP PSF. The glint visible near the lower edge of the first panel is not modeled at this stage. The third panel shows an empirical PSF estimate made by aligning and combining cutouts from images of a different target, HR3188, which did not saturate. All panels have been scaled linearly from their 5th to 99.5th percentile values.}
  \end{figure}

Perhaps the most difficult part of this process is to identify a suitable model to be used as a matched filter for cross-correlation. Using the as-designed phase map for the vAPP coronagraph in Clio, as well as a simple model of the Magellan pupil and the POPPY physical optics package\cite{POPPY}, we generated the simulated vAPP PSF shown in the middle panel of Figure~\ref{fig:psf-model-comparison}. The orientation of the twin vAPP PSFs was measured by Otten \emph{et al.} in Ref. \citenum{Otten2017} and attempts to vary rotation angle did not substantially improve our fit, so the simulated PSFs are generated for $\pm 26^\circ$ rotation only.

These models did not incorporate any uncorrected residual wavefront error, but this did not appear to impair our ability to register two images to each other using their displacements from a common model PSF image. Extending the physical optics model to reproduce not only the PSF morphology but the effects of imperfect AO correction (i.e. Strehl ratio $< 1.0$) remains a challenge to be tackled in future work.

\section{CONCLUSIONS}
\label{sec:conclusions}

Despite the challenges posed by the intense brightness of Sirius, we have developed algorithms that are robust enough to align and background-subtract Clio+vAPP data for use in coronagraphic post-processing for exoplanet searches. The combination of heuristics and iterative algorithms for robustness gives us a suite of software that will generalize to the analysis of other Clio+vAPP data. Furthermore, a better understanding of the unique vAPP three-part PSF structure and its behavior in real data will prove invaluable for developing software to analyze images from vAPP coronagraphs installed in MagAO-X.

%%%%%%%%%%%%%%%%%%%%%%%%%%%%%%%%%%%%%%%%%%%%%%%%%%%%%%%%%%%%%
\acknowledgments     %>>>> equivalent to \section*{ACKNOWLEDGMENTS}

Support for JRM to conduct these observations was provided, in part, under contract with the California Institute of Technology (Caltech)/Jet Propulsion Laboratory (JPL) funded by NASA through the Sagan Fellowship Program executed by the NASA Exoplanet Science Institute. KMM's and LMC's work is supported by the NASA Exoplanets Research Program (XRP) by cooperative agreement NNX16AD44G.

This research was supported in part by NSF MRI Award \#1625441 (MagAO-X).

This work made use of POPPY, an open-source optical propagation Python package originally developed for the James Webb Space Telescope project\cite{POPPY}, as well as Astropy, a community-developed core Python package for Astronomy\cite{Astropy}. Our project also acknowledges the assistance of Asher Haug-Baltzell and his contributions to FINDR\cite{Haug-Baltzell2016}, software which enables the use of cloud computing infrastructure to distribute computing workloads.

%%%%%%%%%%%%%%%%%%%%%%%%%%%%%%%%%%%%%%%%%%%%%%%%%%%%%%%%%%%%%
%%%%% References %%%%%

\bibliography{report}   %>>>> bibliography data in report.bib
\bibliographystyle{spiebib}   %>>>> makes bibtex use spiebib.bst

\end{document}